\documentclass[journal]{IEEEtran}
\usepackage{bm}
\usepackage{mathrsfs}
\usepackage[dvips]{graphicx}
\usepackage{cite}
\usepackage{color}
\usepackage{amsmath}
\usepackage[noend]{algpseudocode}
\usepackage{algorithmicx,algorithm}
\usepackage{subfigure}
\usepackage{xpatch}

\makeatother

\begin{document}

\title{DemodNet: Learning Soft Demodulation from Hard Information Using Convolutional Neural Network
\renewcommand{\thefootnote}{\fnsymbol{footnote}}
\thanks{S. Zheng, S. Chen, and X. Yang are with Science and Technology on Communication Information Security Control Laboratory, Jiaxing 314033, China. (e-mail: lianshizheng@126.com, sicanier@sina.com, yxn2117@126.com).}
\thanks{X. Zhou and P. Qi are with the State Key Laboratory of Integrated Service Networks, Xidian University, Xi'an 710071, China (e-mail: zxy0686@126.com, phqi@xidian.edu.cn).}
}
\author{Shilian Zheng,
        Xiaoyu Zhou,
        Shichuan Chen,
        Peihan Qi,
        and Xiaoniu Yang
}

\maketitle
\begin{abstract}
Soft demodulation is a basic module of traditional communication receivers. It converts received symbols into soft bits, that is, log likelihood ratios (LLRs). However, in the non-ideal additive  white Gaussian noise (AWGN) channel, it is difficult to accurately calculate the LLR. In this letter, we propose a demodulator, DemodNet, based on a fully convolutional neural network with variable input and output length. We use hard bit information to train the DemodNet, and we propose log probability ratio (LPR) based on the output layer of the trained DemodNet to realize soft demodulation. The simulation results show that under the AWGN channel, the performance of both hard demodulation and soft demodulation of DemodNet is very close to the traditional methods. In three non-ideal channel scenarios, i.e., the presence of frequency deviation, additive generalized Gaussian noise (AGGN) channel, and Rayleigh fading channel, the performance of channel decoding using the soft information LPR obtained by DemodNet is better than the performance of decoding using the exact LLR calculated under the ideal AWGN assumption.
\end{abstract}

\IEEEpeerreviewmaketitle
\section{Introduction}
\IEEEPARstart{I}{n} a communication system, the transmitted information is channel coded and modulated and then propagated through the channel to the receiver. The receiver uses a synchronization mechanism to obtain symbols, and then recovers the original information through demodulation and channel decoding \cite{1}. Demodulation here plays a crucial role in the information recovery process. Demodulation can be categorized into soft demodulation and hard demodulation. Hard demodulation converts the received symbols into bit streams, while the soft demodulation outputs the level of confidence of each bit according to the received symbols, which is usually expressed in terms of log-likelihood ratio (LLR). The output of the demodulation is often provided to the channel decoder for decoding. In general, the bit error rate (BER) performance of decoding using soft demodulation output is better than that of decoding using hard demodulation output.

With the development of deep learning technology, the application of deep learning in physical layer communication is becoming more and more extensive \cite{2}\cite{3}. There are some works which use neural networks to achieve demodulation. For example, in \cite{4} deep belief network and stacked autoencoder were used for signal demodulation on short-distance multipath channels. In \cite{5} a convolutional neural network (CNN) was used to demodulate the bipolar extended binary phase shift keying signal to solve the problem of serious inter-symbol interference. In \cite{6} CNN was used to realize FSK demodulation under Rayleigh fading channels. All of these works considered the problem of hard demodulation. In terms of soft demodulation, in order to reduce the computational complexity of accurate LLRs, in \cite{7} exact LLRs were used as labels and a neural network was used to approximate these LLRs to implement soft demodulation. This method can approximate exact LLRs with lower computational complexity. However, the method relies on exact LLRs to train the network and the calculation of exact LLRs will have certain difficulties in practical situations. In general, the calculation of LLR generally assumes that the channel is an additive Gaussian white noise (AWGN) channel. In practical communications process, the signal arrived at the receiver may be influenced by non-ideal factors such as radio frequency (RF) impairments, multipath fading and interference. For these complex non-ideal AWGN environment, the accurate calculation of LLR may not be available and the real LLRs may be different from the LLRs calculated under the assumption of AWGN channel. If the LLRs calculated under the assumption of AWGN channel are used as the labels for training the LLRnet, the LLRnet will approximate these deviated LLRs. Therefore, LLRnet cannot solve the problem of soft demodulation when exact LLRs are not available in practical system with non-ideal AWGN channel.

To solve this problem, in this letter, we use hard information as labels to train a CNN with multiple binary classifiers to implement hard demodulation. We refer to this network as DemodNet in this letter. Based on the output layer of the trained DemodNet, we use the logarithm of the ratio of the probability of a bit being category 0 and the probability of a bit being category 1 as the soft information, so as to realize soft demodulation. The obtained soft information can be used for channel decoding. Because DemodNet is trained with hard information bits, in various complex channel environments, we can obtain labeled training samples by transmitting specific information bit streams and collecting the corresponding received signals at the receiving end, thereby avoiding the problem that the exact LLR cannot be accurately calculated. We finally verified the effectiveness of this method by simulation.

The rest of the letter is organized as follows. In Section II, traditional soft demodulation and hard demodulation is discussed. In Section III, the proposed DemodNet is given in detail. In Section IV, simulation results are given. In Section V, we conclude the letter.

\section{System Model}

In the traditional wireless communication system, the information to be transmitted is first channel-coded and then mapped into symbols by the modulator. The symbols are filtered and then propagated through the channel to the receiving end. The receiver first obtains the estimated value of each symbol through symbol synchronization, and then uses the demodulator to realize the demapping of the symbol to the information hard bit or soft bit. The channel decoder uses the demodulation output to recovers the original information bit stream. In order to focus on the demodulation module, we assume that shaping and symbol synchronization are ideal, so that the received symbol can be expressed as:
\begin{equation}
r(n) = h(n) * s(n){e^{j2\pi \Delta fn}} + w(n),
\end{equation}
where $r(n)$ is the received symbol, $s(n)$ is the transmitted symbol, $\Delta f$ is the normalized frequency deviation, and $w(n)$ is the noise. The difference of the local oscillators' frequency of the transmitter and receiver and the Doppler effect caused by the relative motion will cause the carrier frequency deviation of the received signal.
In this letter, we consider two distribution of noise $w(n)$: AWGN and additive generalized Gaussian noise (AGGN), the probability distribution function of which is
\begin{equation}
f(w) = \frac{\rho }{{2\gamma \Gamma (1/\rho )}}\exp \left\{ { - {{\left| {\frac{{w - \mu }}{\gamma}} \right|}^\rho }} \right\},
\end{equation}
where $\mu$ is the mean, $\rho$ is the shape parameter and $\Gamma ( \cdot )$ is Gamma function. As for the channel response $h(n)$, we consider two scenarios, ideal channel and multipath fading channel. When ideal channel is considered, $h(n)$ is simply an ideal impulse function $\delta (n)$. As for multipath fading channel, Rayleigh fading is considered where the envelop of $h(n)$ follows Rayleigh distribution.

As pointed out earlier, demodulation can be categorized into soft demodulation and hard demodulation. We give a brief introduction of the two demodulations in the following.

\subsection{Soft Demodulation}
For a symbol with $M$ bits, the LLR is defined as the logarithm of the ratio of the probability that a bit takes the value 0 and the probability that a bit takes the value 1\cite{8}, i.e.,
\begin{equation}
{\zeta _i} \buildrel \Delta \over = \log \left( {\frac{{\Pr \left\{ {{b_i} = 0|r} \right\}}}{{\Pr \left\{ {{b_i} = 1|r} \right\}}}} \right),i = 1,2,...,M.
\end{equation}
With an AWGN channel assumption, the exact computation of LLR expression (which is denoted as ExactLLR hereafter) is
\begin{equation}
\begin{array}{l}
{\zeta _i} = \log \left( {\frac{{\sum\nolimits_{s \in {\cal S}_i^0} {\exp \left( { - \frac{{\left\| {r - s} \right\|_2^2}}{{{\sigma ^2}}}} \right)} }}{{\sum\nolimits_{s \in {\cal S}_i^1} {\exp \left( { - \frac{{\left\| {r - s} \right\|_2^2}}{{{\sigma ^2}}}} \right)} }}} \right)\\
 = \log \left( {\frac{{\sum\nolimits_{s \in {\cal S}_i^0} {\exp \left( { - \frac{{{{({\mathop{\rm Re}\nolimits} \{ r\}  - {\mathop{\rm Re}\nolimits} \{ s\} )}^2} + {{({\mathop{\rm Im}\nolimits} \{ r\}  - {\mathop{\rm Im}\nolimits} \{ s\} )}^2}}}{{{\sigma ^2}}}} \right)} }}{{\sum\nolimits_{s \in {\cal S}_i^1} {\exp \left( { - \frac{{{{({\mathop{\rm Re}\nolimits} \{ r\}  - {\mathop{\rm Re}\nolimits} \{ s\} )}^2} + {{({\mathop{\rm Im}\nolimits} \{ r\}  - {\mathop{\rm Im}\nolimits} \{ s\} )}^2}}}{{{\sigma ^2}}}} \right)} }}} \right),\\
i = 1,2,...,M,
\end{array}
\end{equation}
where ${\mathop{\rm Re}\nolimits} \left\{  \cdot  \right\}$ and ${\mathop{\rm Im}\nolimits} \left\{  \cdot  \right\}$ denote taking the real and imaginary part respectively, ${\sigma ^2}$ is the noise variance, ${\cal S}_i^d$ is the subset of symbols or constellation points for which the $i$-th bit is equal to $d \in \{ 0,1\}$. Because exponential calculation and logarithmic calculation are complicated, log max is usually used in practical systems to obtain an approximate LLR \cite{9}.

It should be pointed out that (4) gives the LLR expression under the assumption of AWGN. For other complex channels that are difficult to model theoretically, it is difficult to obtain the exact LLR expression. In this case, if we still use (4) to calculate the LLR, there may be a large deviation from the unknown true LLR, which may affect the performance of subsequent channel decoding.

\subsection{Hard Demodulation}
Different from the purpose of soft demodulation, hard demodulation wishes to obtain the result of bit decision. Its purpose is to make the recovered bits and the actual transmitted bits as identical as possible, thereby reducing the bit error rate of demodulation. One method is to decide based on LLR as
\begin{equation}
{\widehat b_i} = \left\{ {\begin{array}{*{20}{c}}
{0,}\\
{1,}
\end{array}} \right.\begin{array}{*{20}{c}}
{if}\\
{if}
\end{array}\begin{array}{*{20}{c}}
{{\zeta _i} > 0,}\\
{{\zeta _i} \le 0.}
\end{array}
\end{equation}

\section{The Proposed DemodNet}
\subsection{The Structure of DemodNet}
\begin{figure*}[!t]
\centering
\includegraphics[width=16cm]{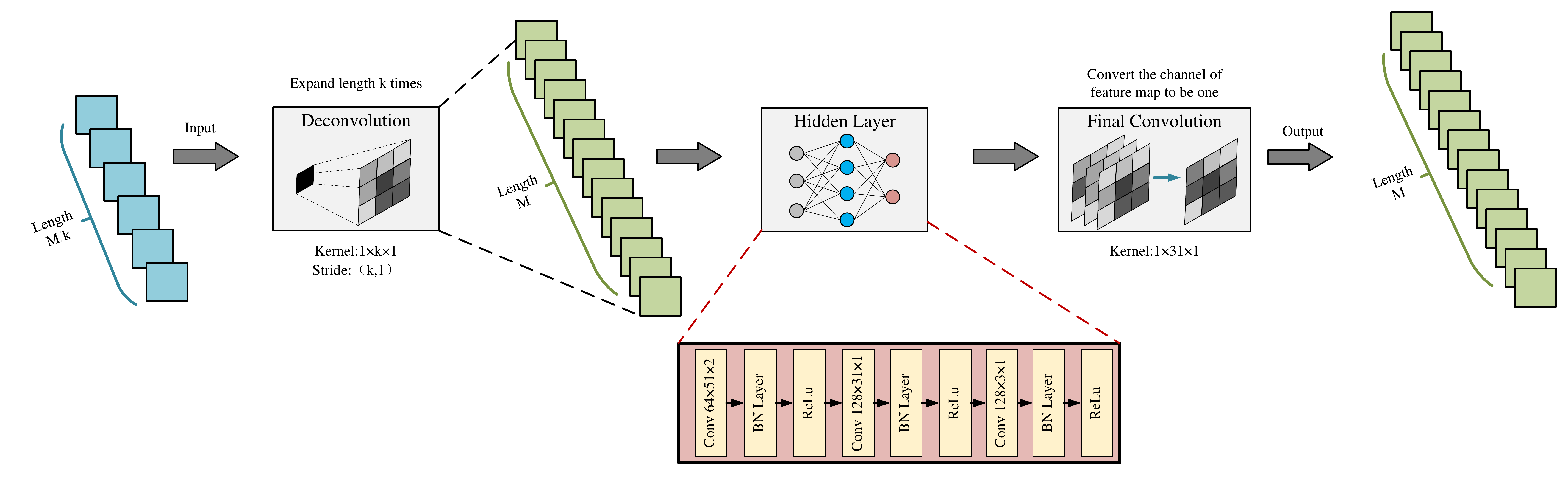}
\caption{Structure of DemodNet. In this figure, the length of input symbol sequence is $M/k$, the ``Deconvolution'' layer is used to expand the length of input signal by $k$ times, the ``Hidden Layer'' is composed of three convolutional layers and there are Batch Normalization layer (``BN Layer'') and activation layer (``ReLU'') between convolutional layers. The ``Final Convolution'' is a convolution layer with one kernel and the kernel size is $31 \times 1$. The length of the final output is the same as the output of ``Deconvolution'' layer.}
\label{Fig.1}
\end{figure*}
We treat the hard demodulation problem as an $M$ binary classification problem. The label of each classifier is 0 or 1. We design DemodNet as shown in Fig. 1 to solve the classification problem. In practice, the number of the received symbols is not fixed, and the number of bits output by the demodulation is also different. In order to make the trained network adapt to the demodulation tasks with different number of symbols, the output length of the network must change with the number of input symbols, so we design a fully convolutional network with variable input and output length to achieve hard demodulation. For multi-ary modulations, the length of the demodulated output bit stream is generally greater than the number of input symbols, so we first deconvolve the input at the first layer of the network to make the output length equal to the number of demodulated bits. The expanded signal undergoes three-layer convolutions to extract the features. Traditional CNNs \cite{10} usually use a fully connected layer to obtain a fixed-length vector after convolution, but doing so will make the network output fixed and cannot be adjusted according to the network input. Unlike this, we replace the fully connected layer with a convolution layer with only one convolution kernel at the last layer of the network, thereby ensuring that the output length is $M$, which is $k$ times the input length, where $k$ is the number of bits of a symbol. Finally, we use Sigmoid activation function for binary classification. The Sigmoid expression is
\begin{equation}
f\left( {{z_i}} \right) = \frac{1}{{1 + \exp \left( { - {z_i}} \right)}},
\end{equation}
where $z_i$ is the input of Sigmoid which also known as logits. In summary, the DemodNet has the following two characteristics. First, the network can accept input of any length without requiring all data to have the same size. Second, the network uses deconvolution for upsampling, so that the network can be used for demodulation tasks of different modulation types.

DemodNet can be trained based on the training set. The training set contains the received signal samples, i.e., symbol sequence after synchronous, and the corresponding label, i.e., the transmitted bit sequence, which can be expressed as
\begin{equation}
{\cal D} = \left\{ {\left( {{{\left[ {{\mathop{\rm Re}\nolimits} \left( \bm r \right),{\mathop{\rm Im}\nolimits} \left( \bm r \right)} \right]}^{\left( i \right)}},{\bm b^{\left( i \right)}}} \right)} \right\}_{i = 1}^N,
\end{equation}
where $\bm r=[r(1),r(2),...,r(\frac{M}{k})]^T$, $\bm b=[b_1,b_2,...,b_M ]^T$ is the transmitted bit sequence, $N$ is the number of samples in the training set. The training of DemodNet uses the sum of binary cross entropy as the loss function. For a sample of bit length $M$, the loss can be represented as
\begin{equation}
{\cal L} =  - \sum\limits_{i = 1}^M {\left( {{y_i}\log {{\widehat y}_i} + \left( {1 - {y_i}} \right)\log \left( {1 - {{\widehat y}_i}} \right)} \right)} ,
\end{equation}
where $y_i$ is the true label of the $i$-th bit, ${\widehat y}_i$ is Sigmoid output of the corresponding bit. We use Adam optimizer \cite{11} to minimize the loss $\cal L$.
\subsection{Soft Information of DemodNet}
Although the learning of DemodNet is based on hard bit information as a label, it also provides a soft demodulation mechanism. As we know, Sigmoid output ${\widehat y}_i$ can represent the probability that the bit is 1 in a certain sense, therefore, we calculate the following soft information for soft demodulation
\begin{equation}
{\xi _i} = \log \left( {\frac{{1 - {{\widehat y}_i}}}{{{{\widehat y}_i}}}} \right),i = 1,2,...,M,
\end{equation}
In addition, if we have got the value of the logits, then
\begin{equation}
{\xi _i} =  - {z_i},i = 1,2,...,M.
\end{equation}
We refer to ${\xi _i}$ as log probability ratio (LPR) in the rest of the letter and the obtained LPR can be used for the subsequent soft decision channel decoding. As a result, DemodNet, which learns based on hard bit information, also has the capability of ``soft demodulation''.

\section{Simulation Results}
\subsection{Performance under AWGN Channel}
We first consider the performance of hard demodulation. Five modulations are considered in the simulation: BPSK, QPSK, 16QAM, 64QAM, and 256QAM, all assuming an ideal AWGN channel. In the training set, the Eb/N0 of the BPSK and QPSK signals range from 0 dB to 8 dB with an interval of 1 dB. The Eb/N0 of 16QAM, 64QAM, and 256QAM ranges from 0 dB to 12 dB, 0 dB to 16 dB, and 0 dB to 20 dB, respectively. For each modulation, the number of samples per Eb/N0 is 100,000. In each test data set, for each modulation, the number of samples per Eb/N0 is 50,000. Each sample contains 100 symbols. DemodNet is trained separately for each modulation. All parameters of the DemodNet are initialized randomly with Gaussian distribution. During training, the mini-batch size is 128 and the maximum number of epochs is 15. The initial learning rate is 0.003, and after every 3 epochs, the learning rate drops to 1/2 of the previous learning rate. Fig. 2 shows the simulation results. It can be seen that the BER performance of DemodNet is very close to the ideal hard demodulation under the five modulations, which shows the effectiveness of DemodNet to implement hard demodulation.
\begin{figure}[!t]
\centering
\includegraphics[width=2.5in]{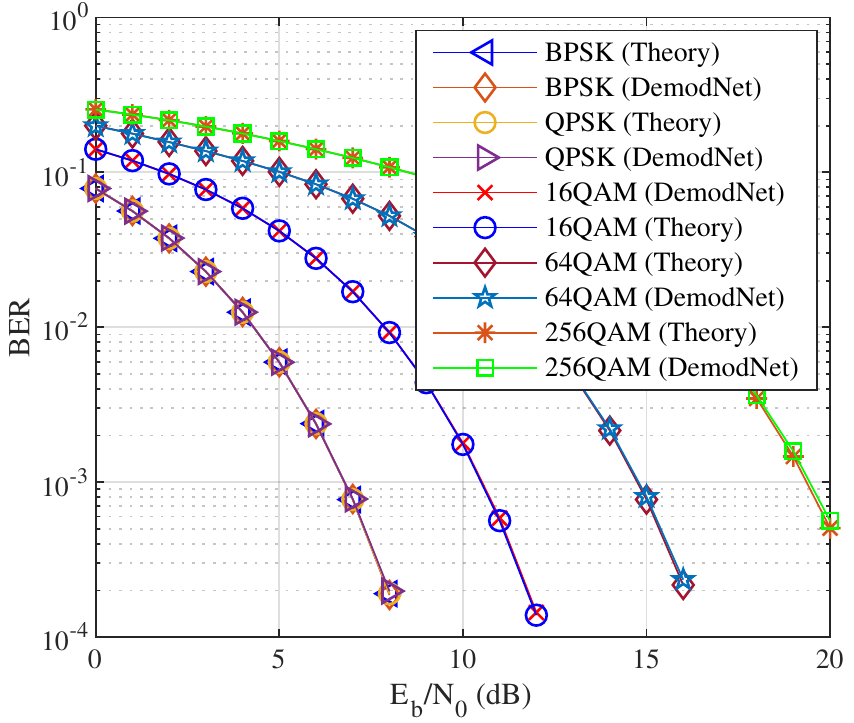}
\caption{Hard demodulation performance of DemodNet. Under the five modulations, the BER performance of DemodNet for hard demodulation almost coincides with the theoretical results.}
\label{Fig.2}
\end{figure}

Next, we analyze the performance of the LPR soft information learned by DemodNet. In the simulation, the channel coding used is a convolutional code with coding rate 1/2, constraint length 7, and code generator polynomials 171 and 133. The soft demodulation output is provided to a Viterbi decoder \cite{12} whose traceback depth is 32 for soft decoding. Fig. 3 shows the simulation results. It can be seen that under BPSK and 16QAM, the BER performance of Viterbi decoding using LPR obtained by DemodNet is very close to that of Viterbi decoding using ExactLLR. In the case of 64QAM, the performance of LPR is slightly worse than that of ExactLLR. These results show that although DemodNet has never seen ExactLLR, from the perspective of supporting Viterbi decoding, it also obtains demodulation performance close to that of ExactLLR, which validates its effectiveness for soft demodulation.
\begin{figure}[!t]
\centering
\includegraphics[width=2.5in]{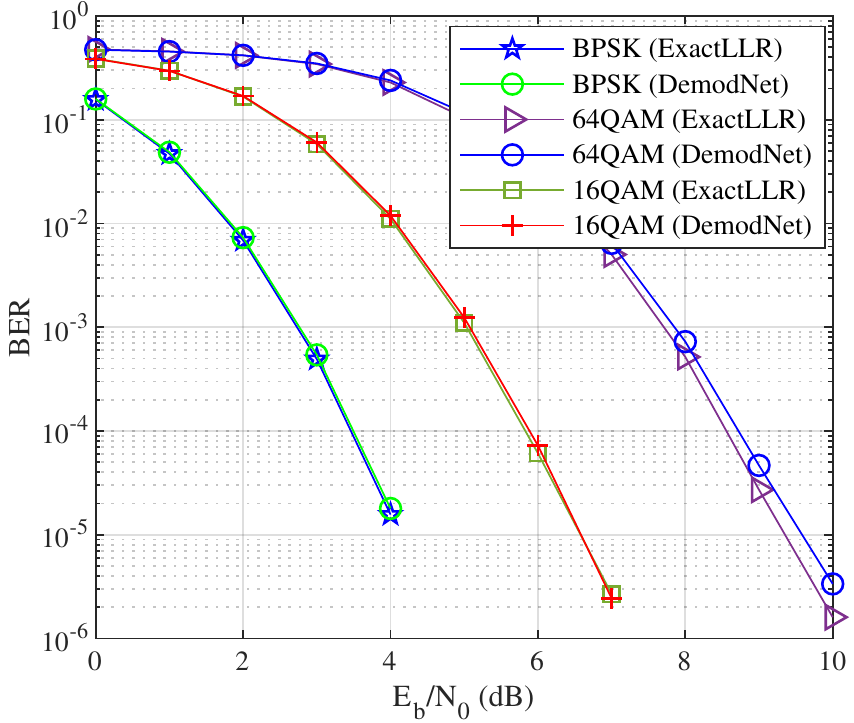}
\caption{BER performance of Viterbi decoding using LPR and LLR in AWGN channel.}
\label{Fig.3}
\end{figure}

\subsection{Effects of Frequency Deviation}
We now consider the performance with frequency deviation in AWGN channel. Fig. 4 shows the simulation results, where the frequency deviation is 0.005 times the symbol rate. The channel coding used is a convolutional code as mentioned before. Similarly, the BER in the figure refers to the BER output by Viterbi soft decision. It can be seen that due to the frequency deviation, the performance of the ExactLLR calculated according to the ideal AWGN channel deteriorates significantly. The performance of the LPR learned by DemodNet is better than the exactLLR, which shows the superiority of DemodNet.
\begin{figure}[!t]
\centering
\includegraphics[width=2.5in]{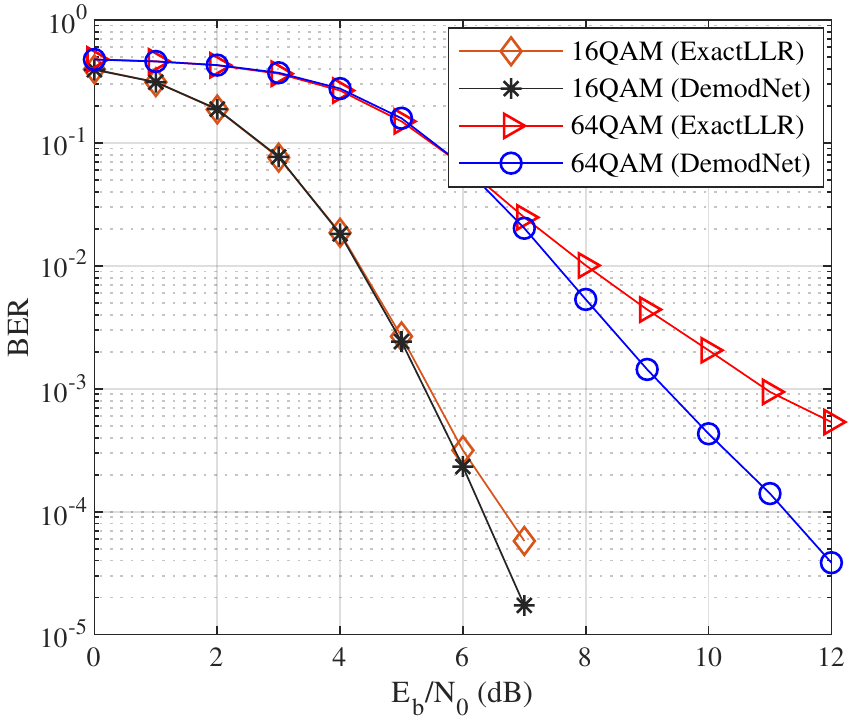}
\caption{Soft demodulation performance under frequency deviation.}
\label{Fig.4}
\end{figure}
\begin{figure}[!t]
\centering
\subfigure[]
{\includegraphics[width=2.5in]{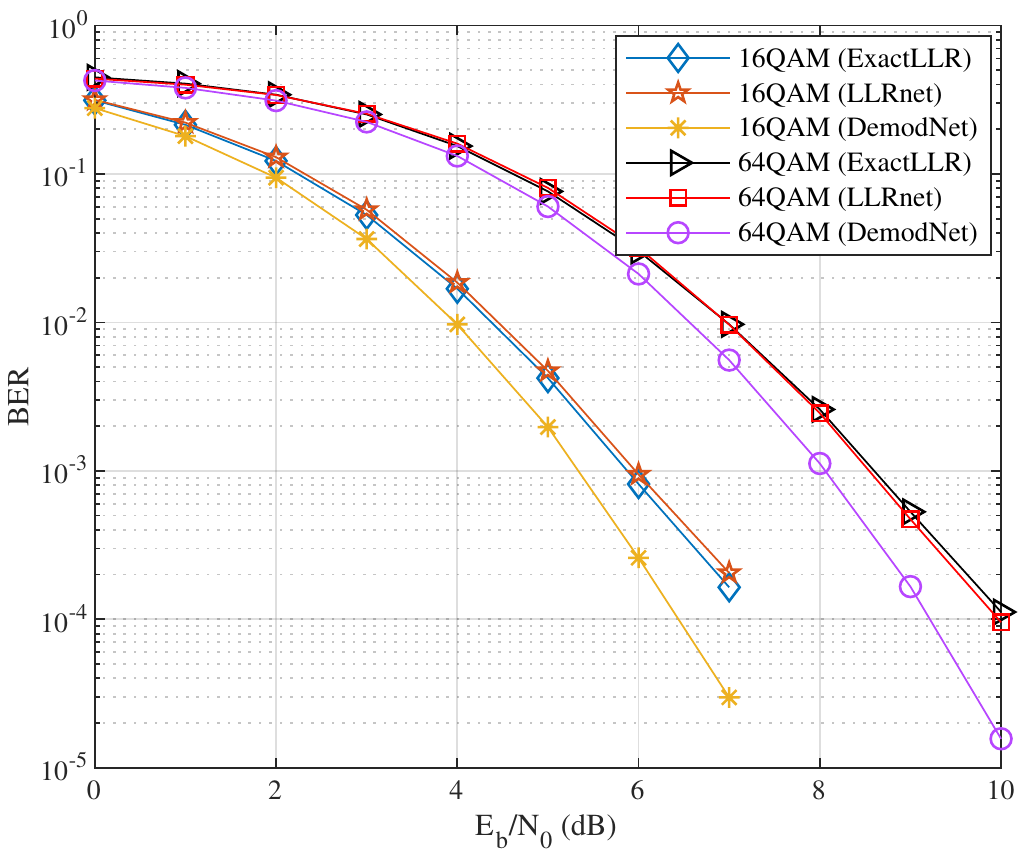}}
\hfil
\subfigure[]
{\includegraphics[width=2.5in]{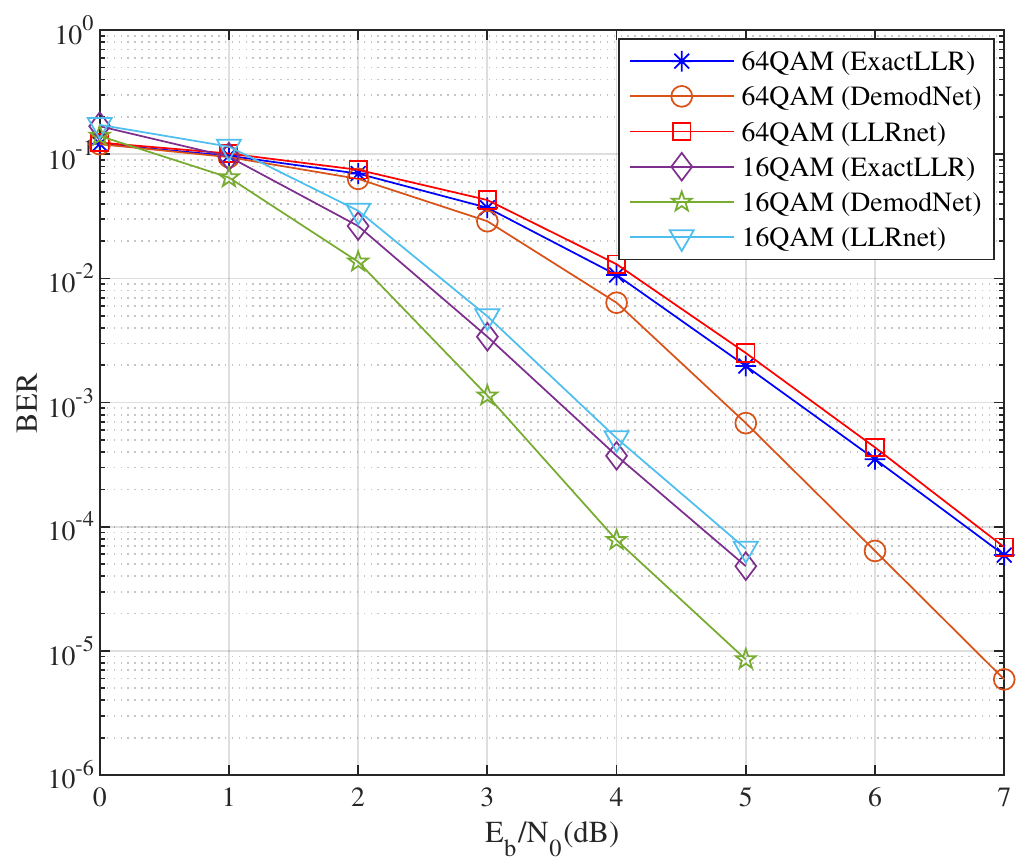}}
\caption{ Soft demodulation performance in AGGN channel. Two different channel coding are used: (a) convolutional code and (b) turbo code.}
\label{Fig.5}
\end{figure}

\subsection{Performance under AGGN Channel}
We then give the simulation results under the AGGN channel. The AGGN parameters are set to: $\mu  = 0,\gamma  = 1,\rho  = 1.$  We assume that the demodulation does not know the real distribution of the channel and thus ExactLLRs can only be obtained under the AWGN assumption. Fig. 5 shows the BER results of channel decoding using the soft demodulation information. Two channel coding methods are considered, convolutional code as mentioned before and turbo code with code rate 1/3. It can be seen that with both channel coding methods under AGGN, the performance of the ExactLLR is degraded, which is obvious because the ExactLLR is obtained under the AWGN assumption and they do not necessarily match the AGGN channel. Because DemodNet directly uses the signal data in the case of the AGGN channel for training, the performance of the decoding using DemodNet LPR is better than that of the decoding using ExactLLR.

Furthermore, we compare the performance of DemodNet with that of LLRnet in [7], in which a deep neural network (DNN) was used to approximate LLRs for reducing computational complexity. The LLRnet can only approximate exactLLRs obtained under AWGN. The BER curve of LLRNet is very close to the BER curve of ExactLLR and it performs worse than DemodNet under AGGN noise, just as Fig. 5 (a) and Fig. 5 (b) shows.

\subsection{Performance under Rayleigh Fading Channel}
We perform a simulation to evaluate the performance of DemodNet under frequency flat Rayleigh fading channel. The simulation assumes the symbol rate is 1Msps, and the maximum Doppler shift is set to 30. Adaptive equalization with least mean square (LMS) algorithm is adopted to equalize the signals. The number of taps for LMS is 5 and the step size is 1/10 of the estimated maximum allowable step size. For equalization methods, a known training sequence is needed to adjust tap weights. So, we add a fixed 500-symbol sequence in front of the channel encoded symbol stream. The BER results of decoding under Rayleigh fading channel is shown in Fig. 6. 16QAM and 64QAM are valuated and the previously mentioned convolutional code is used. We can see that the performance of DemodNet is better than that of ExactLLR which further shows the superiority of our method in Rayleigh fading channel.
\begin{figure}[!t]
\centering
\includegraphics[width=2.5in]{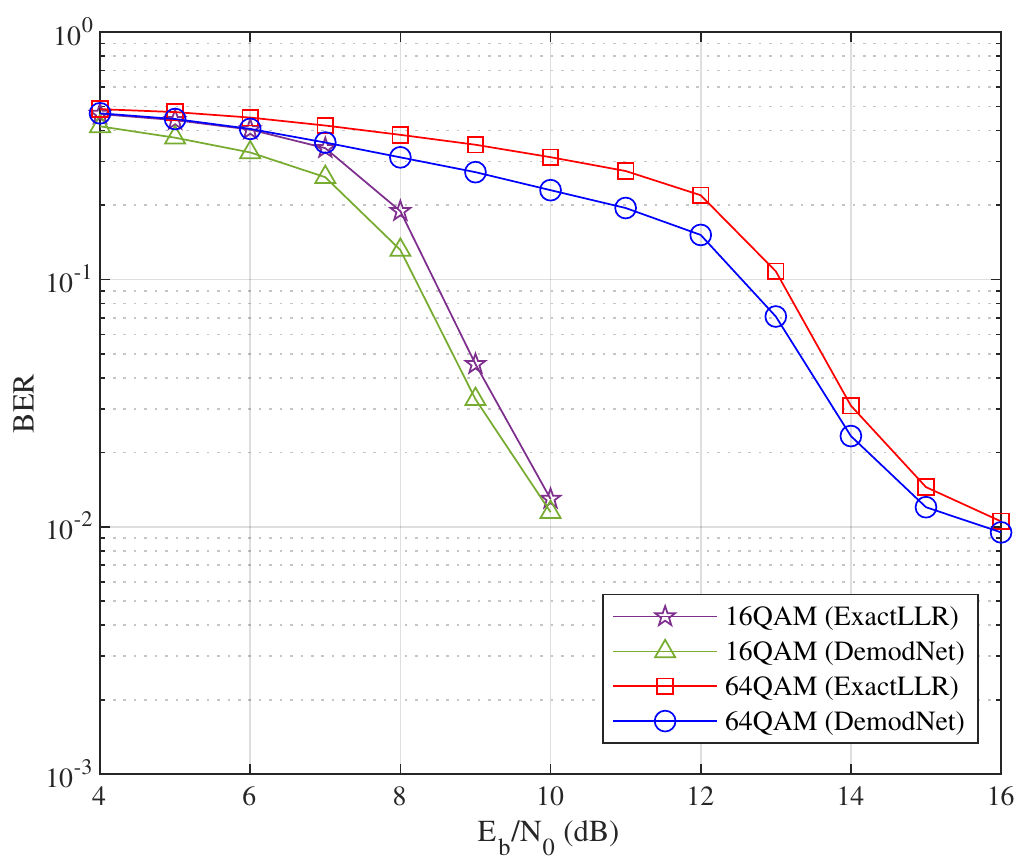}
\caption{Soft demodulation performance under Rayleigh fading channel.}
\label{Fig.6}
\end{figure}

\subsection{Complexity Analysis}
In DemodNet, we assume the input of each layer is ${h_{in}} \times {w_{in}} \times {c_{in}}$, and the corresponding output is ${h_{out}} \times {w_{out}} \times {c_{out}}$, then the computational complexity of convolutional layer and deconvolution layer is
\begin{equation}
{C_{conv}} \sim O({c_{in}} \times {h_{out}} \times {w_{out}} \times {c_{out}} \times k),
\end{equation}
where $k$ is the size of convolution kernel, ${c_{in}}$ is the channel of input, and ${c_{out}}$ is the channel of output, i.e., the number of convolution kernels. The computational complexity of the batch normalization layer and the ReLU layer are both
\begin{equation}
{C_{bn}} \sim O({h_{in}} \times {w_{in}} \times {c_{in}}).
\end{equation}
For the demodulation of 64QAM, give a $M$-symbol sequence, the total computational complexity of DenodNet include 4636044$M$ multiplication, 4636044$M$ addition and 1920$M$ comparator. The computational complexity of ExactLLR include 198$M$ multiplication, 564$M$ addition and 70$M$ exponential or logarithm calculation. It can be seen that the multiplication, addition and addition computational complexity of DenodNet is larger than that of ExactLLR. However, the DenodNet eliminates exponential or logarithm calculation, which require a lot of computation. Furthermore, the computation of the DemodNet can be parallelized and accelerated to reduce the computational time.

\section{Conclusions}
In this letter, we have proposed DemodNet for hard demodulation and soft demodulation. DemodNet can learn soft information from hard information to support subsequent channel decoding. We have verified the performance of DemodNet through simulation experiments. In the ideal AWGN channel, the hard demodulation BER performance of DemodNet is very close to the theoretical value and the BER performance of soft decoding using DemodNet LPR is close to that of decoding using ExactLLR. In the cases of carrier frequency deviation, AGGN channel and Rayleigh fading channel, the decoding performance using DemodNet LPR is superior to the decoding performance using ExactLLR calculated with the assumption of ideal AWGN. Therefore, DemodNet may provide a new way for soft demodulation under complex channel environments where LLRs are difficult to be accurately estimated via analytical methods.


\end{document}